\documentclass{webofc}
\usepackage[varg]{txfonts}   
\newcommand{\trento}{T\raisebox{-.5ex}{R}ENTo}
\begin{document}
\title{Holographic transport coefficients and jet energy loss for the hot and dense quark-gluon plasma}
%
%

\author{\firstname{Joaquin} \lastname{Grefa}\inst{1,2}\fnsep\thanks{\email{vgrefaju@central.uh.edu}} \and
        \firstname{Mauricio} \lastname{Hippert}\inst{3} \and
        \firstname{Raghav} \lastname{Kunnawalkam Elayavalli}\inst{4} \and
        \firstname{Jacquelyn} \lastname{Noronha-Hostler}\inst{2} \and
        \firstname{Israel} \lastname{Portillo}\inst{1} \and
        \firstname{Claudia} \lastname{Ratti}\inst{1} \and
        \firstname{Romulo} \lastname{Rougemont}\inst{5} 
}

\institute{Department of Physics, University of Houston, Houston, Texas 77204, USA 
\and Department of Physics, Kent State University, Kent, Ohio 44243, USA
\and
           Illinois Center for Advanced Studies of the Universe, Department of Physics, University of Illinois at Urbana-Champaign, Urbana, Illinois 61801, USA 
\and
        Department of Physics and Astronomy, Vanderbilt University, Nashville, Tennessee, 37235, USA
\and
           Instituto de F\'{i}sica, Universidade Federal de Goi\'{a}s, Av. Esperan\c{c}a - Campus Samambaia, CEP 74690-900, Goi\^{a}nia, Goi\'{a}s, Brazil
          }

\abstract{We employ an Einstein-Maxwell-dilaton model, based on the gauge/gravity correspondence, to obtain the thermodynamics and transport properties for the hot and dense quark-gluon plasma. The model, which is constrained to reproduce lattice QCD thermodynamics at zero density, predicts a critical point and a first order line at finite temperature and density, is used to quantify jet energy loss through simulations of high-energy collision events.
}
\maketitle

Lattice QCD simulations have predicted the analytical crossover phase transition nature between the confined hadronic gas and the deconfined strongly interacting liquid known as quark-gluon plasma (QGP) at vanishing density \cite{Ratti:2018ksb}. However, at finite density, where the crossover is conjectured to evolve into a first order line with a critical end point (CEP), ab-initio lattice calculations are hindered by the sign problem. Moreover, lattice simulations face a significant challenge related to the computation of transport observables \cite{Meyer:2011gj} which are needed to understand the QGP behavior to perturbatutions near and out of equilibrium. 

In order to explore the QCD phase diagram, where first principle calculations are not possible, we should employ an effective model that agrees with lattice QCD thermodynamics at zero density, exhibits the nearly inviscid fluid behaviour of the QGP and describes deconfined strongly interacting matter. The holographic EMD model from Ref. \cite{Critelli:2017oub}, based on the gauge-gravity duality \cite{Maldacena:1997re,Gubser:1998bc}, fulfills these requirements and provides the thermodynamics for the hot and baryon dense QGP \cite{Grefa:2021qvt} with the advantage of being able to handle near equilibrium calculations to compute several transport coefficients and energy loss observables \cite{Grefa:2022sav}.

\section{Equation of state and baryon transport coefficients at finite temperature and density}
 
\begin{figure}
    \centering
    \includegraphics[width=0.45\textwidth]{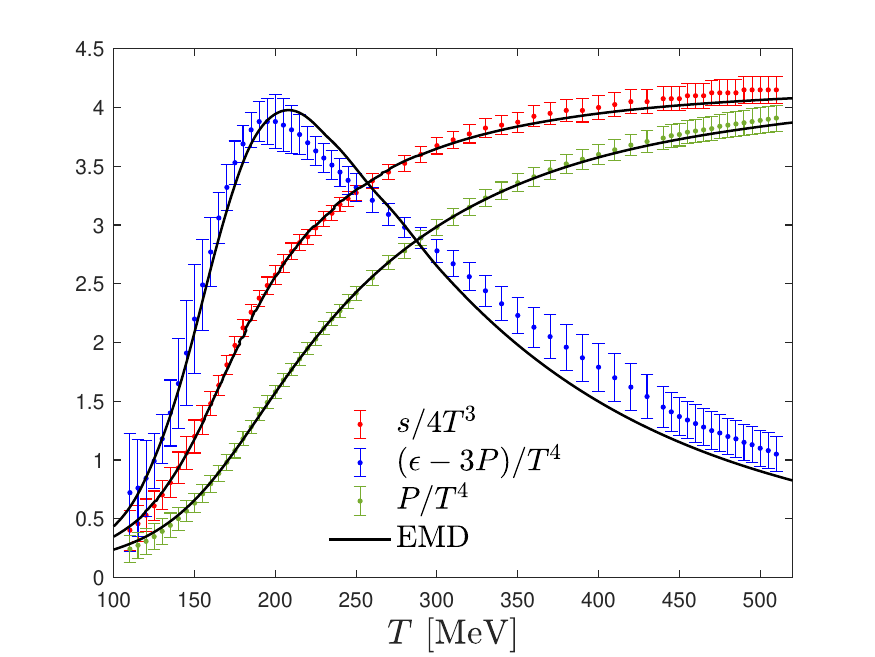}
    \includegraphics[width=0.45\textwidth]{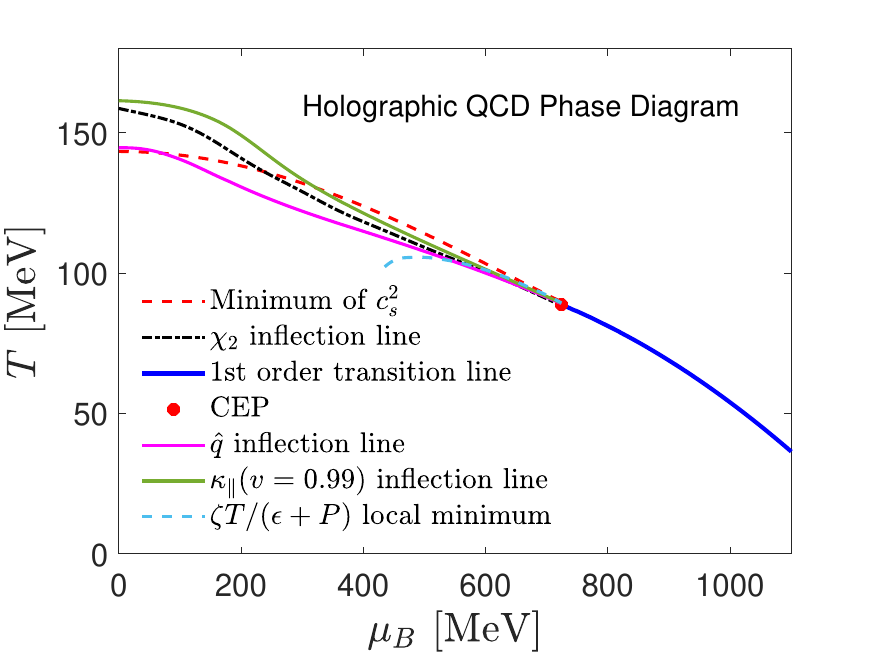}
    \caption{Left: Temperature dependence of the holographic (solid lines) entropy density, pressure and trace anomaly and their corresponding lattice results from \cite{Borsanyi:2013bia}. Right: Holographic QCD phase diagram where the crossover has been characterized by different characteristic points from equilibrium and transport observables (taken from Ref. \cite{Grefa:2022sav}) with all these trajectories converging at the CEP.}
    \vspace{-0.55cm}
    \label{fig:EoS_results}
\end{figure}

 Based on the seminal works of Refs.~\cite{Gubser:2008ny,DeWolfe:2010he, DeWolfe:2011ts}, the non-conformal bottom-up Einstein-Maxwell-dilaton (EMD) holographic model from Ref. \cite{Critelli:2017oub} has been devised to quantitatively describe the physics of the strongly coupled QGP. The construction of the gravitational action includes a 5-dimensional bulk metric $g_{\mu\nu}$ coupled to real scalar field $\phi$ (the dilaton field) responsible of breaking the conformal symmetry of the theory through a potential $V(\phi)$. The effects due to nonzero baryon chemical potential $\mu_{B}$ are taken into account by adding a Maxwell field $A_{\mu}$ coupled to the dilaton field through a function $f(\phi)$ in the bulk action. Since QCD is the target dual gauge theory at the AdS boundary (far away from the black hole horizon in the gravitational theory), the free parameters of the model and the potentials $V(\phi)$ and $f(\phi)$ are fixed by dynamically matching the holographic entropy density and second order baryon susceptibility at $\mu_{B}=0$ with the corresponding lattice QCD results for 2+1 flavors and physical quark masses from Refs.  \cite{Borsanyi:2013bia,Bellwied:2015lba}. The results of this fitting are shown in the left panel of Fig.~\ref{fig:EoS_results}. On the other hand, any observable at finite baryon chemical potential as well as other quantities computed at zero chemical potential constitute predictions of this EMD model, which includes the location of the QCD critical point at $(T,\mu_B)_{\text{CEP}}\approx(89,724)$~MeV and a first order transition line \cite{Critelli:2017oub,Grefa:2021qvt}. Additionally, the EMD thermodynamics shows an excellent quantitative agreement with the state-of-the-art lattice QCD data from Ref.~\cite{Borsanyi:2021sxv}. More recently, by using Bayesian inference techniques, different functional parametrizations of the free functions in the EMD model were investigated and suggest a strong preference for the existence of a QCD CEP besides delimiting its possible location when the model is constrained by lattice QCD data \cite{Hippert:2023bel}.

\label{sec-1}

In addition to the equilibrium properties, the transport coefficients, which describe the QGP response to perturbations near equilibrium and are an input in realistic hydrodynamical simulations, were calculated for the baryon charge case in Ref. \cite{Grefa:2022sav} by employing the same EMD model of Refs. \cite{Critelli:2017oub,Grefa:2021qvt} across the phase diagram including the CEP and transition line. The set of computed transport coefficients include the baryon and thermal conductivities, baryon diffusion, heavy quark drag force, Langevin diffusion coefficients, jet quenching parameter, and bulk and shear viscosities. The description of the holographic QCD phase diagram was extended by including not only the characteristic points from equilibrium variables, but also the behaviour of transport observables in Ref.~\cite{Grefa:2022sav}. This is shown in the right panel of Fig.~\ref{fig:EoS_results} where the relevant characteristic trajectories of the equation of state and transport variables, corresponding to extrema or inflection points, converge at the CEP.

\section{Jet energy loss}

The transport coefficients, particularly the jet quenching parameter $\hat{q}$, obtained from the EMD model of Refs. \cite{Critelli:2017oub,Grefa:2021qvt}, offer a unique opportunity for modeling parton energy loss since the holographic equation of state displays an excellent agreement with lattice QCD data. Thus, we are able to perform simulations of high energy heavy ion collisions in a self-consistent way by incorporating the equilibrium variables and transport coefficients from the same model. 

The jet energy loss produced by the interaction with the QGP and characterized by the temperature/time dependent jet quenching parameter is the final topic of this section. In order to include quenching effects, we compute the time an emission within a jet takes to behave as an independent source of radiation. This formation time can be obtained from \cite{Apolinario:2022vzg}: 
\begin{equation}
    \tau_{form}=\frac{1}{2Ez(1-z)(1-\cos{\theta})},
\end{equation}
where $E$ is the energy of the incoming particle that splits into a pair with an opening angle $\theta$, each carrying a fraction $z$ and $(1-z)$ of the parent parton.
Then, at each formation time, the emission energy and angle are computed from a distribution that depends on $\hat{q}$. Such a distribution reads:  
\begin{equation}    P(\theta,\omega)=\alpha\omega\theta^{3}\sqrt{\frac{2\omega}{\hat{q}}}L\exp{\left(\frac{-\theta^{2}\omega^{2}}{\sqrt{2\omega\hat{q}}}\right)},
\end{equation}
where $\alpha$ is a constant, $\omega$ is the emission energy, $\theta$ is the emission angle, and $L$ is the path length. We translate the temperature dependence of the holographic $\hat{q}$ to a time dependence by employing \trento \,\cite{Moreland:2014oya} to model the initial state of a relativistic heavy ion collision. With the holographic equation of sate as an input, \trento \,yields a map of the energy density in the transverse plane (perpendicular to the collision axis) just after two heavy ions collide. This energy density map is then evolved in time by using ideal Bjorken hydrodynamics to model the QGP. For the toy model presented in this work, we took the average of all the events within $0.25 - 0.30$ centrality with JETSCAPE parameters. 

\begin{figure}
    \centering
    \includegraphics[width=0.45\textwidth]{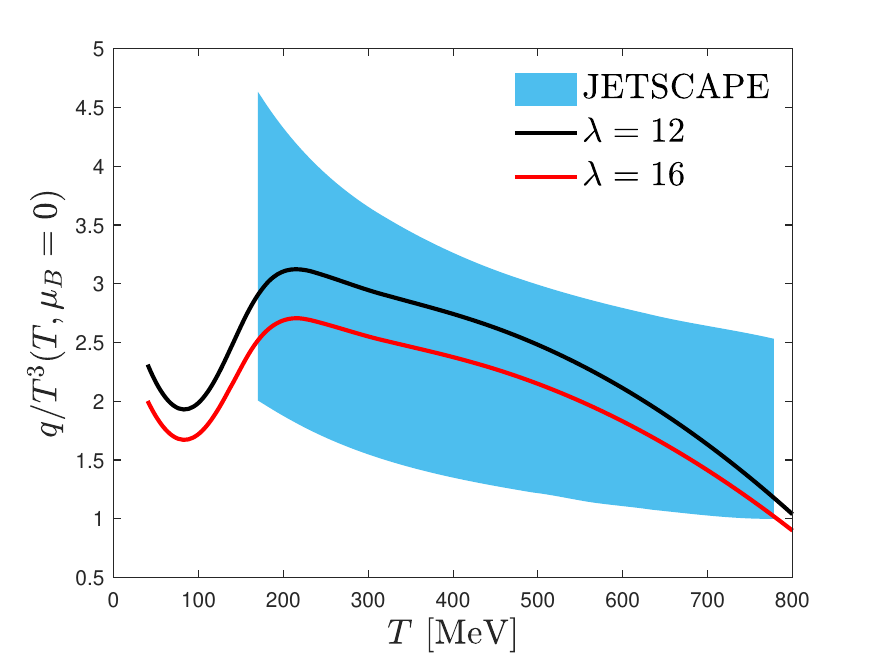}
    \includegraphics[width=0.45\textwidth]{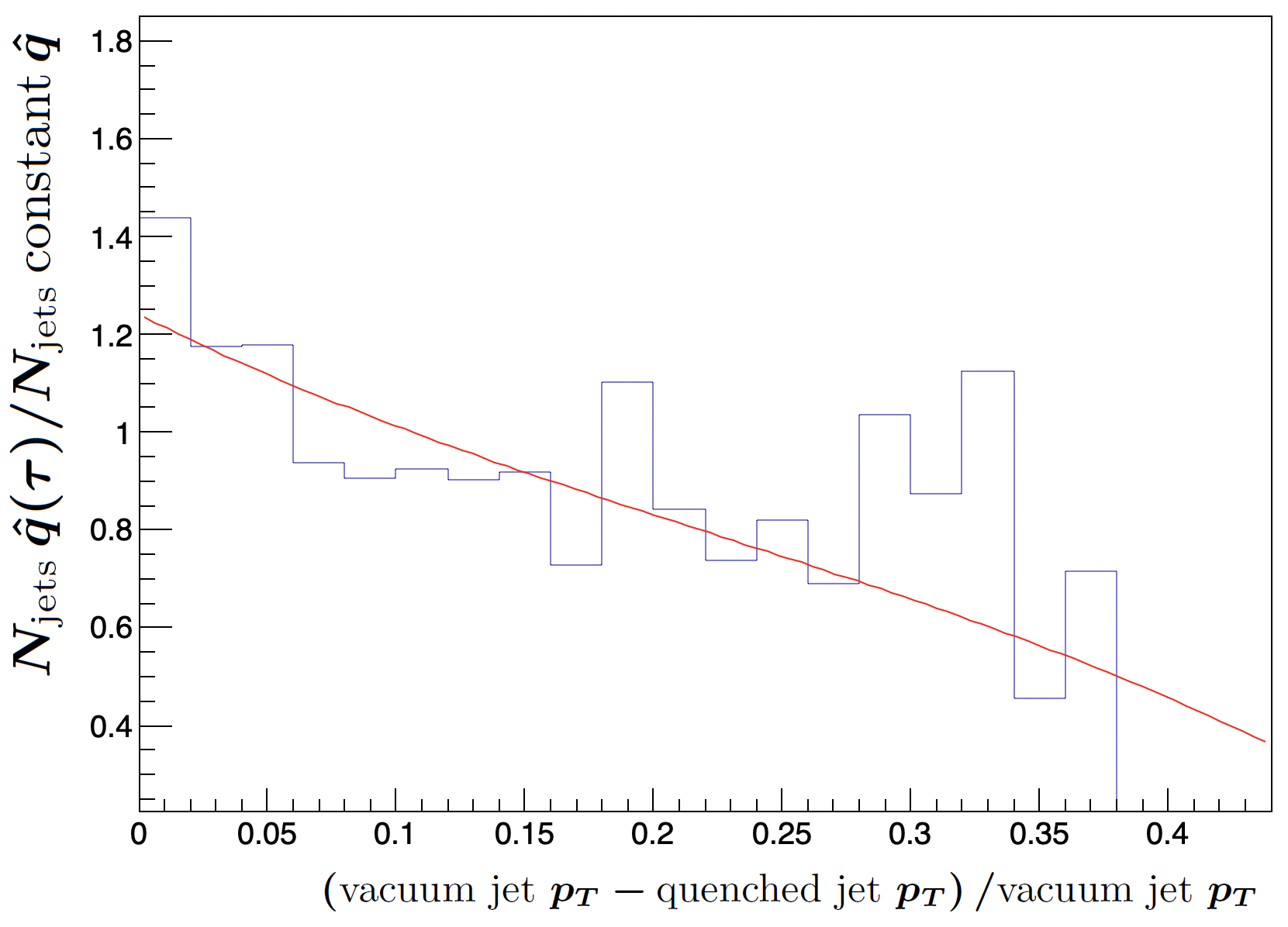}
    \caption{Left: holographic jet quenching parameter $\hat{q}$ as a function of the temperature for two values of the t'Hooft parameter $\lambda$, and its comparison with the JETSCAPE results taken from Ref. \cite{Apolinario:2022vzg}. Right: jet quenching as a function of normalized transverse momentum $p_{T}$ described as a ratio of the number of jets with a time dependent $\hat{q}(\tau)$ over number of jets with a constant $\hat{q}$ }
    \vspace{-0.55cm}
    \label{fig:jet_quenching}
\end{figure}

The jet quenching parameter $\hat{q}$ is computed by considering a probe NambuGoto (NG) action for a classical string on top of the solutions for the EMD fields \cite{Grefa:2022sav}. The NG action is proportional to the t'Hooft coupling ($\sqrt{\lambda}$) which in our case is a free parameter that should be fixed phenomenologically. We compare the holographic $\hat{q}$ to the corresponding result from JETSCAPE summarized in Ref. \cite{Apolinario:2022vzg} to constraint the value of the t'Hooft coupling. This is show in the left panel of Fig. \ref{fig:jet_quenching} for two different values of $\lambda$. For our simulations performed in PYTHIA, we fixed this value to be $\lambda=16$. In the right panel of Fig.~\ref{fig:jet_quenching}, we show on the vertical axis the ratio of the number of jets when the jet quenching parameter is a function of time $\hat{q}=\hat{q}(\tau)$ over the number of jets when $\hat{q}$ has a constant value of 1 $\text{GeV}^2/\text{fm}$. The horizontal axis is the normalized transverse momentum where the vacuum jet $p_{T}$ corresponds to the absence of quenching. The downward trend represented by the red curve indicates that there are more jets that lose less energy (near the region of zero to small normalized transverse momentum) compared to the number of jets that are heavily quenched (to the right of the horizontal axis). This is reasonable since one may expect jets to be heavily quenched in the initial stages of the collision where the temperature is the highest and the energy density is large in comparison with a reduced transverse momentum suppression as the system expands and cools down, a process not necessarily described by the constant $\hat{q}$ scenario. 

In conclusion, in this work we have shown how a time dependent $\hat{q}$ is relevant to model jet energy loss, which motivates a more profound  study of parton energy loss as a function of time. Besides improving the statistics, we plan to refine our toy model by including the transport coefficients in the Bjorken expansion, and a more realistic hydrodynamic simulation.

\textbf{ Acknowledgements:} This material is based upon work supported by the National Science Foundation under grants No. PHY-2208724 and No. PHY-2116686 and in part by the U.S. Department of Energy, Office of Science, Office of Nuclear Physics, under Award Number DE-SC0022023, DE-SC0023861. This work was supported in part by the National Science Foundation (NSF) within the framework of the MUSES collaboration, under grant number No. OAC-2103680. R.R. acknowledges financial support by National Council for Scientific and Technological Development (CNPq) under grant number 407162/2023-2. RKE acknowledges startup funds from Vanderbilt University which supported JG and MH to travel to Vanderbilt in the summer of 2023 where a part of this work was done.
%
%
%
\bibliography{references.bib}

\end{document}